\documentclass[cameraready]{Interspeech}
\usepackage{threeparttable}
\usepackage{multirow}
\usepackage{booktabs,multirow,}
\usepackage{cite}
\usepackage{url}

\title{Language-Invariant Multilingual Speaker Verification for the TidyVoice 2026 Challenge}

\author[affiliation={1,4}]{Ze}{Li}
\author[affiliation={4}]{Xiaoxiao}{Miao}
\author[affiliation={2}]{Juan}{Liu}
\author[affiliation={2,3,4}, orcid=0000-0002-6406-1983, correspondingauthor]{Ming}{Li}

\address{
    $^1$ School of Computer Science, Wuhan University, Wuhan, China \\
    $^2$ School of Artificial Intelligence, Wuhan University, Wuhan, China \\
    $^3$ School of Artificial Intelligence, The Chinese University of Hong Kong, Shenzhen, China \\
    $^4$ Digital Innovation Research Center, Duke Kunshan University, Kunshan, China
}

\email{lize389@whu.edu.cn, xiaoxiao.miao@dukekunshan.edu.cn, liujuan@whu.edu.cn, ming.li.cuhksz@gmail.com}


\usepackage{comment}

\begin{document}

\maketitle

\begin{abstract}
    Multilingual speaker verification (SV) remains challenging due to limited cross-lingual data and language-dependent information in speaker embeddings. This paper presents a language-invariant multilingual SV system for the TidyVoice 2026 Challenge. We adopt the multilingual self-supervised w2v-BERT 2.0 model as the backbone, enhanced with Layer Adapters and Multi-scale Feature Aggregation to better exploit multi-layer representations. A language-adversarial training strategy with a Gradient Reversal Layer is applied to promote language-invariant speaker embeddings. Moreover, a multilingual zero-shot text-to-speech system is used to synthesize speech in multiple languages, improving language diversity. Experimental results demonstrate that fine-tuning the large-scale pretrained model yields competitive performance, while language-adversarial training further enhances robustness. In addition, synthetic speech augmentation provides additional gains under limited training data conditions. Source code is available at \url{https://github.com/ZXHY-82/LI-MSV-TidyVoice2026}.
\end{abstract}

\section{Introduction}
Speaker verification (SV) aims to verify the identity of speakers by analyzing their voice samples. In recent years, with the rapid development of deep neural networks and the availability of large-scale labeled speech datasets, deep learning-based SV systems~\cite{cai_resnet, ecapa-tdnn, cam++, redimnet} have achieved remarkable performance across a wide range of acoustic conditions. However, the performance of SV systems degrades significantly under the language mismatch condition, which is further exacerbated by the field’s reliance on English-centric datasets. In addition, the limited availability of multilingual speech from individual speakers often leads to speaker embeddings that entangle identity with language-specific characteristics, reducing cross-lingual generalization and robustness.

Large-scale self-supervised Pre-Trained Models (PTMs), such as WavLM~\cite{wavlm}, wav2vec 2.0~\cite{wav2vec2.0}, HuBERT~\cite{hubert} and w2v-BERT 2.0~\cite{w2v-bert-2.0}, are trained on hundreds of thousands or even millions of hours of unlabeled speech data and provide rich speech representations. These models have been increasingly adopted in research to enhance performance on various downstream tasks, including the SV task. In particular, the w2v-BERT 2.0 PTM is trained on 4.5 million hours of unlabeled speech spanning 143 languages, making it highly suitable for cross-lingual SV tasks. Li et al.~\cite{li2025enhancing} built a state-of-the-art SV system based on w2v-BERT 2.0. In their approach, Layer Adapters~\cite{cai_asr_sv} are applied to the outputs of each Conformer layer to reduce dimensionality and facilitate domain adaptation. The adapted features from all layers are then aggregated using a Multi-scale Feature Aggregation (MFA)~\cite{mfa} framework to generate speaker embeddings, and Low-Rank Adaptation (LoRA)~\cite{lora} is employed during training to fine-tune the model efficiently.

Speaker embeddings extracted from multilingual data often contain not only identity information but also language-specific characteristics, which can reduce cross-lingual generalization and robustness. To mitigate this issue, it is necessary to encourage the learning of language-invariant speaker representations. Similar approaches have been adopted in other scenarios, such as age-invariant speaker representation learning~\cite{cross-age}, where adversarial training is used to suppress age-related information in the embeddings.

In this work, we follow the approach in~\cite{li2025enhancing} to build SV systems based on the w2v-BERT 2.0 PTM. To further improve cross-lingual robustness, a language-adversarial training strategy is introduced to encourage the learning of language-invariant speaker representations and suppress language-related variations in the embedding space. In addition, although the official TidyVoiceX training set is multilingual, each speaker typically contains only two to three languages, which limits language diversity for robust cross-lingual speaker modeling. Recent advances in Zero-Shot Text-To-Speech (ZS-TTS)~\cite{cosyvoice3, qwen3-tts} technology have made it possible to synthesize high-quality speech that preserves the vocal characteristics of a target speaker using only a few seconds of reference audio, without requiring speaker-specific training. This enables flexible voice cloning across different languages and textual contents. Leveraging this capability, we adopt a multilingual ZS-TTS system, Qwen3-TTS~\cite{qwen3-tts}, to synthesize speech in additional languages for each speaker. We aim to enrich the multilingual speech data for each speaker, thereby facilitating the learning of more language-invariant speaker representations.

\begin{figure*}
  \centering
  \includegraphics[width=0.9\linewidth]{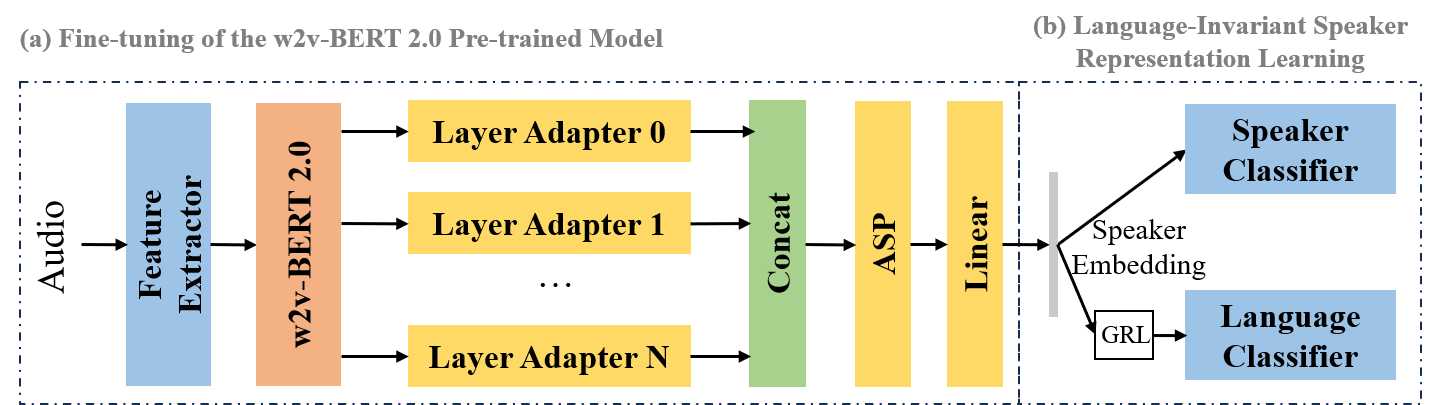}
  \caption{Overview of the w2v-BERT 2.0-based speaker verification system with language-invariant learning}
  \label{fig:system_framework}
\end{figure*}

\section{Methods}
\subsection{Fine-tuning of the w2v-BERT 2.0 Pre-trained Model}
W2v-BERT 2.0 is a large-scale multilingual self-supervised speech representation model, developed as part of the SeamlessM4T framework~\cite{w2v-bert-2.0}. It extends the original w2v-BERT architecture~\cite{w2v-bert} by employing a 24-layer Conformer encoder and jointly optimizing contrastive and masked prediction objectives. The model is trained on approximately 4.5 million hours of unlabeled speech data across 143 languages, enabling it to learn rich and robust multilingual speech representations.

Our approach follows the method of~\cite{li2025enhancing}. As shown in Fig.\ref{fig:system_framework}(a), given an input utterance $x$, we first extract Fbank features and feed them into the pre-trained w2v-BERT 2.0 model. The hidden representations $h_i$ from each Conformer layer are then passed through individual Layer Adapters~\cite{cai_asr_sv}, which reduce feature dimensionality and facilitate domain adaptation for the speaker verification task. The adapted features $h^\prime_i$ from all layers are concatenated and subsequently aggregated using an attentive statistics pooling (ASP)~\cite{asp_pooling} module, followed by a linear projection layer to obtain the final speaker embedding $e$. During training, we employ Low-Rank Adaptation (LoRA)~\cite{lora} to efficiently fine-tune the pre-trained model.
\begin{equation}
    [h_0,h_1, \ldots,h_L] = \text{w2v-BERT-2.0}(\text{Fbank($x$)})
\end{equation}
\begin{equation}
    h^\prime_i = \text{Layer Adapter}_i(h_i), \quad i = 0, 1, \ldots, L
\end{equation}
\begin{equation}
    e = \text{Linear}(\text{ASP}(\text{Concat}(h^\prime_0,h^\prime_1,\ldots,h^\prime_L)))
\end{equation}

\subsection{Language-Invariant Speaker Representation Learning}
Language mismatch introduces undesired variability into speaker embeddings, potentially degrading the performance of speaker verification systems in multilingual scenarios. To address this issue, we aim to learn language-invariant speaker representations by explicitly removing language-related information as shown in Fig.\ref{fig:system_framework}(b).

Specifically, we introduce an auxiliary language classifier connected to the speaker embedding extractor through a Gradient Reversal Layer (GRL)~\cite{GRL}. Given the extracted speaker embedding $e$, the language classifier predicts the language label, while the GRL reverses the gradient during backpropagation. This adversarial learning strategy encourages the embedding extractor to produce representations that are discriminative for speaker identity while being uninformative about language identity. The final loss is formulated as:
\begin{equation}
\mathcal{L}_\text{spk}(e) = l_\text{spk}\big(C_\text{spk}(e), y_\text{spk}\big)
\end{equation}
\begin{equation}
\mathcal{L}_\text{lang}(e) = l_\text{lang}\Big(C_\text{lang}\big(\text{GRL}_{\lambda_\text{GRL}}(e)\big), y_\text{lang}\Big)
\end{equation}
\begin{equation}
\mathcal{L}_\text{total} = \mathcal{L}_\text{spk} + \lambda_\text{lang} \, \mathcal{L}_\text{lang}
\end{equation}

where $C_\text{spk}(\cdot)$ and $C_\text{lang}(\cdot)$ are the speaker and language classifiers, $l_\text{spk}(\cdot)$ and $l_\text{lang}(\cdot)$ are the corresponding loss functions, $\text{GRL}_{\lambda_\text{GRL}}(\cdot)$ is the gradient reversal layer with scale $\lambda_\text{GRL}$, $\lambda_\text{lang}$ is a hyperparameter controlling the weight of the language loss, and $y_\text{spk}$ and $y_\text{lang}$ are the labels of the speaker and language, respectively.

\begin{figure}
  \centering
  \includegraphics[width=\linewidth]{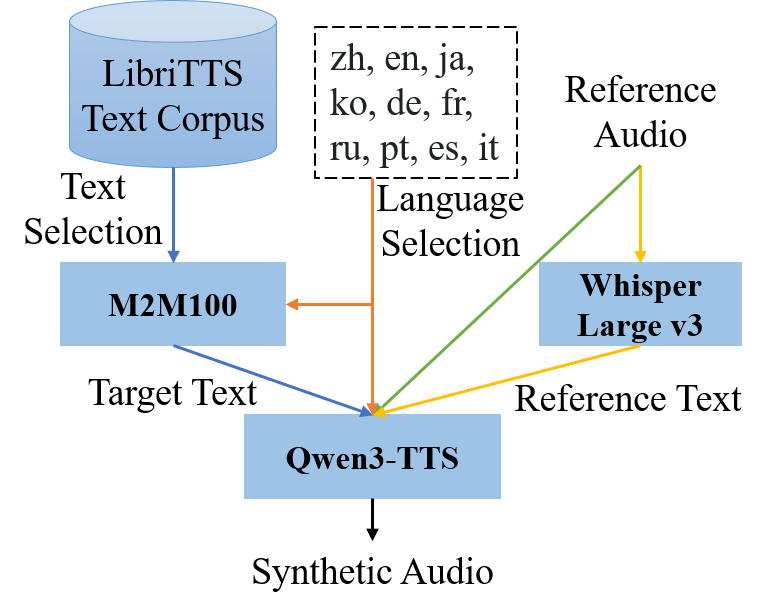}
  \caption{Speech Synthesis Pipeline}
  \label{fig:syn}
\end{figure}

\subsection{Multilingual Synthetic Speech Data Augmentation}

Recent advances in ZS-TTS have enabled high-quality speech synthesis across multiple languages from only a few seconds of reference audio. Leveraging this capability, we generate synthetic speech for each speaker in multiple languages, aiming to improve the language-invariance of speaker embeddings and enhance cross-lingual speaker verification performance.

In this work, we employ Qwen3-TTS~\cite{qwen3-tts} for synthetic speech generation. Qwen3-TTS is a model capable of multilingual voice cloning, enabling it to generate speech in ten languages, including Chinese, English, Japanese, Korean, German, French, Russian, Portuguese, Spanish, and Italian. As shown in Fig.\ref{fig:syn}, for the text corpus, we use English sentences from LibriTTS~\cite{libritts}, which are then translated into the target languages using the M2M100~\cite{m2m100} multilingual translation model. Reference audio is processed with Whisper-large-v3~\cite{whisper-large-v3} to obtain the corresponding transcript, and both the reference audio and text are fed into Qwen3-TTS to generate synthetic speech in the specified language and target text.

\section{Experimental Setup}
\subsection{Datasets}
To train a robust multilingual SV system, in addition to the official TidyVoiceX training set provided by the TidyVoice 2026 Challenge~\cite{tidyvoice2026}, we also incorporate several publicly available datasets, including VoxCeleb2~\cite{vox2dev}, VoxBlink2~\cite{voxblink2}, 3D-Speaker~\cite{3d-speaker}, KeSpeech~\cite{kespeech}, and CN-Celeb1\&2~\cite{cnceleb1, cnceleb2}, to increase speaker and language diversity. The SV performance is evaluated on the official development and evaluation sets provided by the challenge. The evaluation set is further divided into two subsets: tv26\_eval-A, which contains languages seen in the official training set, and tv26\_eval-U, which consists of 38 unseen languages.

\subsection{Multilingual Synthetic Speech Generation}
We use the Qwen3-TTS-12Hz-1.7B-Base~\cite{qwen3-tts} model to perform multilingual voice cloning. From the TidyVoiceX training set, we select up to ten utterances longer than 3 seconds for each speaker as reference audio, resulting in a total of 3,495 reference audios. For each reference audio, we synthesize speech in the ten languages supported by Qwen3-TTS, generating ten utterances per language. In total, 349,500 synthetic utterances are generated for multilingual data augmentation.

\subsection{Training Details}
Our training process is divided into two stages, including (i) Large-Scale Speaker Model Pre-training and (ii) Fine-tuning on TidyVoiceX Training Set with Language-Invariant Learning.
\subsubsection{Large-Scale Speaker Model Pre-training}
In this stage, we fine-tune the pre-trained w2v-BERT 2.0 model using several large-scale public speaker datasets, including VoxCeleb2, VoxBlink2, 3D-Speaker, KeSpeech, and CN-Celeb1\&2, to obtain a robust large-scale pre-trained speaker model.

During the initial training phase, the w2v-BERT 2.0 parameters are frozen. The input acoustic features are 80-dimensional fbank coefficients with a frame length of 25ms and a hop size of 10ms. Mean and variance normalization are applied before feeding the features into the model. On-the-fly data augmentation~\cite{on-the-fly} is applied by adding background noise or convolutional reverberation noise. The MUSAN~\cite{musan} and RIR Noise~\cite{RIR} datasets are used as noise sources and room impulse response functions, respectively. AdamW~\cite{adamw} optimizer with weight decay of 1e-4 is used, along with a StepLR scheduler with 5 epochs decay. A linear warm-up schedule is applied during the first 5 epochs to stabilize training, followed by a StepLR scheduler with a decay factor of 0.1, which decreases the learning rate from 1e-4 to 1e-5. ArcFace~\cite{arcface} loss is adopted as the speaker classification objective, with the margin and scale set to 0.2 and 32, respectively. The input frame length is randomly sampled between 200 and 300 frames.

After convergence, the w2v-BERT 2.0 parameters are unfrozen for further fine-tuning. In this phase, the learning rate starts from 1e-5 and gradually decreases to 5e-6 using a cosine decay schedule over 2 epochs, with a total of 4 epochs dedicated to fine-tuning.

\subsubsection{Fine-tuning on TidyVoiceX Training Set with Language-Invariant Learning}
In this stage, the TidyVoiceX training set is incorporated into the training process for further domain adaptation, and a language-adversarial learning strategy is introduced to encourage language-invariant speaker representation learning. Specifically, a language classifier consisting of two linear layers is newly added on top of the speaker embedding. To ensure stable language supervision, all other modules are first frozen, and only the language classifier is trained until convergence. Subsequently, the speaker encoder and other modules are unfrozen, and a GRL is introduced between the speaker embedding and the language classifier. The GRL reverses the gradient from the language classification objective during backpropagation, forcing the encoder to suppress language-related information while preserving speaker-discriminative features. The coefficients $\lambda_{GRL}$ and $\lambda_{lang}$ are both set to 0.1 to control the strength of adversarial learning. The model is trained using a cosine decay learning rate schedule, where the learning rate starts from 1e-5 and gradually decreases to 5e-6 over 2 epochs, with a total of 4 training epochs. The same data augmentation strategies as in the previous stage are applied.
ArcFace loss is adopted as the default loss function due to its wide adoption. For comparison, we also conduct several experiments using the SphereFace2 (SF2)~\cite{sf2} loss function in this stage.

\subsection{Score Calibration}
To further improve score reliability, we adopt a Quality Measure Function (QMF)~\cite{qmf1} to calibrate the SV scores. QMF aims to compensate for score variability caused by differences in speech duration, signal quality, and embedding reliability. The QMF model is trained using trials randomly generated from the TidyVoiceX training set. As described in~\cite{qmf},we adopt the following QMF qualities set $q$ to calibrate the scores:
\begin{equation}
    \mathbf{q} = [\log(d_e),\log(d_t),\lVert e \rVert,\lVert t \rVert,\text{SNR}_e,\text{SNR}_t,s]
\end{equation}

where $\log(d_e)$ and $\log(d_t)$ denote the logarithms of the enrollment and test utterance durations, $\lVert e \rVert$ and $\lVert t \rVert$ represent the magnitudes of the enrollment and test embeddings, $\text{SNR}_e$ and $\text{SNR}_t$ indicate the signal-to-noise ratios of the enrollment and test utterances, and $s$ is the verification score. Only $\text{SNR}_e$ and $\text{SNR}_t$ adopt the Max-Min normalization.

Logistic Regression is adopted to train the QMF model on the generated trials:
\begin{equation}
    \mathbf{s'} = \sigma(\mathbf{w}^\top \mathbf{q} + b)
\end{equation}

where $w$ and $b$ are the learned parameters, and $s'$ is the calibrated score.

During inference, the trained QMF model is applied to produce the final verification score.

\section{Results}
\begin{table*}[htbp]\centering 
    \scriptsize 
    \caption{Performance of the w2v-BERT 2.0 based SV systems on the TidyVoice 2026 development and evaluation sets. The w2v-BERT 2.0\_Based, w2v-BERT 2.0\_Based$_{SF2-A}$, and w2v-BERT 2.0\_Based$_{SF2-C}$ denote models trained with the default ArcFace loss, SphereFace2-A loss, and SphereFace2-C loss, respectively.}
    \tabcolsep=0.6em
     \label{tab:res}
\begin{threeparttable}
    \begin{tabular}{lcccccccc}
    \toprule

    \multirow{2}*{\textbf{Model}} &\multirow{2}*{\textbf{Pretraining Data}} &\multirow{2}*{\textbf{Fine-tuning Data}} & \multicolumn{2}{c}{\textbf{tv26\_dev}} & \multicolumn{2}{c}{\textbf{tv26\_eval-A}} & \multicolumn{2}{c}{\textbf{tv26\_eval-U}} \\
    \cmidrule(lr){4-5} \cmidrule(lr){6-7} \cmidrule(lr){8-9} & & & \textbf{EER(\%)} & \textbf{mDCF$_{0.01}$} & \textbf{EER(\%)} & \textbf{mDCF$_{0.01}$} & \textbf{EER(\%)} & \textbf{mDCF$_{0.01}$} \\
    \midrule
    Official Baseline~\cite{tidyvoice2026} & VoxBlink2 + VoxCeleb2 & TidyVoiceX Train & 3.07 & 0.82 & 9.058 & 0.65 & 11.59 & 0.60 \\
    \midrule
    w2v-BERT 2.0\_Based & \multirow{13}{*}{\begin{tabular}{@{}l@{}}VoxBlink2 \\ + VoxCeleb2 \\ + 3D-Speaker \\ + KeSpeech \\ + CnCeleb1\&2 \end{tabular}} & None & 2.740 & 0.79 & - & - & - & - \\
    \cline{1-1} \cline{3-9}
    w2v-BERT 2.0\_Based &  & \multirow{3}{*}{\begin{tabular}{@{}l@{}}Pretraining Data + \\ TidyVoiceX Train \end{tabular}} & 1.466 & 0.66 & - & - & - & - \\
    w2v-BERT 2.0\_Based$_{SF2-A}$ & & & 1.089 & 0.63 & - & - & - & - \\
    w2v-BERT 2.0\_Based$_{SF2-C}$ & & & 1.065 & 0.62 & 3.061 & 0.24 & \textbf{4.338} & \textbf{0.29} \\
    \cline{1-1} \cline{3-9}
    w2v-BERT 2.0\_Based &  & \multirow{4}{*}{\begin{tabular}{@{}l@{}}TidyVoiceX Train \end{tabular}} & 1.191 & 0.63 & - & - & - & - \\
    w2v-BERT 2.0\_Based$_{SF2-A}$ & &  & 0.966 & 0.61 & - & - & - & - \\
    w2v-BERT 2.0\_Based$_{SF2-C}$ & &  & 0.950 & 0.61 & - & - & - & - \\
    \hspace{1em}+GRL &  &  & 0.937 & 0.60 & 2.964 & 0.23 & 5.020 & 0.30 \\
    \hspace{1em}++QMF &  &  & \textbf{0.893} & \textbf{0.60} & \textbf{2.458} & \textbf{0.21} & 4.451 & 0.29 \\
    \cline{1-1} \cline{3-9}
    w2v-BERT 2.0\_Based$_{SF2-C}$ &  & All Synthetic Data & 1.022 & 0.60 & - & - & - & - \\
    \cline{1-1} \cline{3-9}
    \multirow{2}{*}{w2v-BERT 2.0\_Based$_{SF2-C}$} &  & \multirow{2}{*}{\begin{tabular}{@{}l@{}}TidyVoiceX Train + \\ All Synthetic Data \end{tabular}} & \multirow{2}{*}{0.999} & \multirow{2}{*}{0.61} & \multirow{2}{*}{-} & \multirow{2}{*}{-} & \multirow{2}{*}{-} & \multirow{2}{*}{-} \\
    &  &  &  &  &  &  &  &  \\
    \cline{1-1} \cline{3-9}
    \multirow{2}{*}{w2v-BERT 2.0\_Based$_{SF2-C}$} &  & \multirow{2}{*}{\begin{tabular}{@{}l@{}}TidyVoiceX Train + \\ Sub Synthetic Data \end{tabular}} & \multirow{2}{*}{0.954} & \multirow{2}{*}{0.61} & \multirow{2}{*}{-} & \multirow{2}{*}{-} & \multirow{2}{*}{-} & \multirow{2}{*}{-} \\
    &  &  &  &  &  &  &  &  \\
    \bottomrule
    \end{tabular}
    
\end{threeparttable}
\end{table*}

Table \ref{tab:res} presents the performance of the w2v-BERT 2.0-based SV systems on the TidyVoice 2026 development and evaluation sets. Compared with the official baseline system, SimAM-ResNet34~\cite{simam, tidyvoice2026}, which is also pre-trained on a large amount of data, the SV systems that fine-tune the pre-trained w2v-BERT 2.0 model demonstrate a clear advantage. Even without using the TidyVoiceX training set, the model fine-tuned solely on pre-training data achieves an EER of 2.74\% on the tv26\_dev set, yielding an 11\% relative reduction in EER compared to the baseline value of 3.07\%.

Furthermore, we also explored the effectiveness of different loss functions and fine-tuning data configurations. For the loss function, in addition to the widely used ArcFace loss, we also evaluated SphereFace2 losses with A and C configurations. The results show that models fine-tuned with SphereFace2 significantly outperform those using ArcFace, with SphereFace2-C achieving the best performance. This is mainly attributed to the fact that, unlike ArcFace, which formulates SV as a multi-class classification problem, SphereFace2 adopts a binary classification objective in the hyperspherical space. Since both training and evaluation in SV rely on pairwise similarity comparison, this formulation effectively reduces the mismatch between the training objective and the evaluation protocol.

Regarding the fine-tuning data strategy, we compared using only the TidyVoiceX training set with mixing it with the large-scale pre-training datasets. The results indicate that fine-tuning solely on the TidyVoiceX training set achieves better performance on tv26\_dev and tv26\_eval-A, where the languages are seen during training, but performs worse on tv26\_eval-U, which contains unseen languages. 
This is mainly attributed to the difference between domain specialization and generalization. Fine-tuning exclusively on the TidyVoiceX training set allows the model to focus more on the target domain, leading to better performance on tv26\_dev and tv26\_eval-A, where the languages and acoustic conditions are similar to those seen during training. In contrast, incorporating large-scale multilingual pre-training data improves the model's generalization by exposing it to more diverse languages and acoustic variations. Moreover, the pre-training data may include languages that overlap with those in tv26\_eval-U.

\begin{figure}
  \centering
  \includegraphics[width=\linewidth]{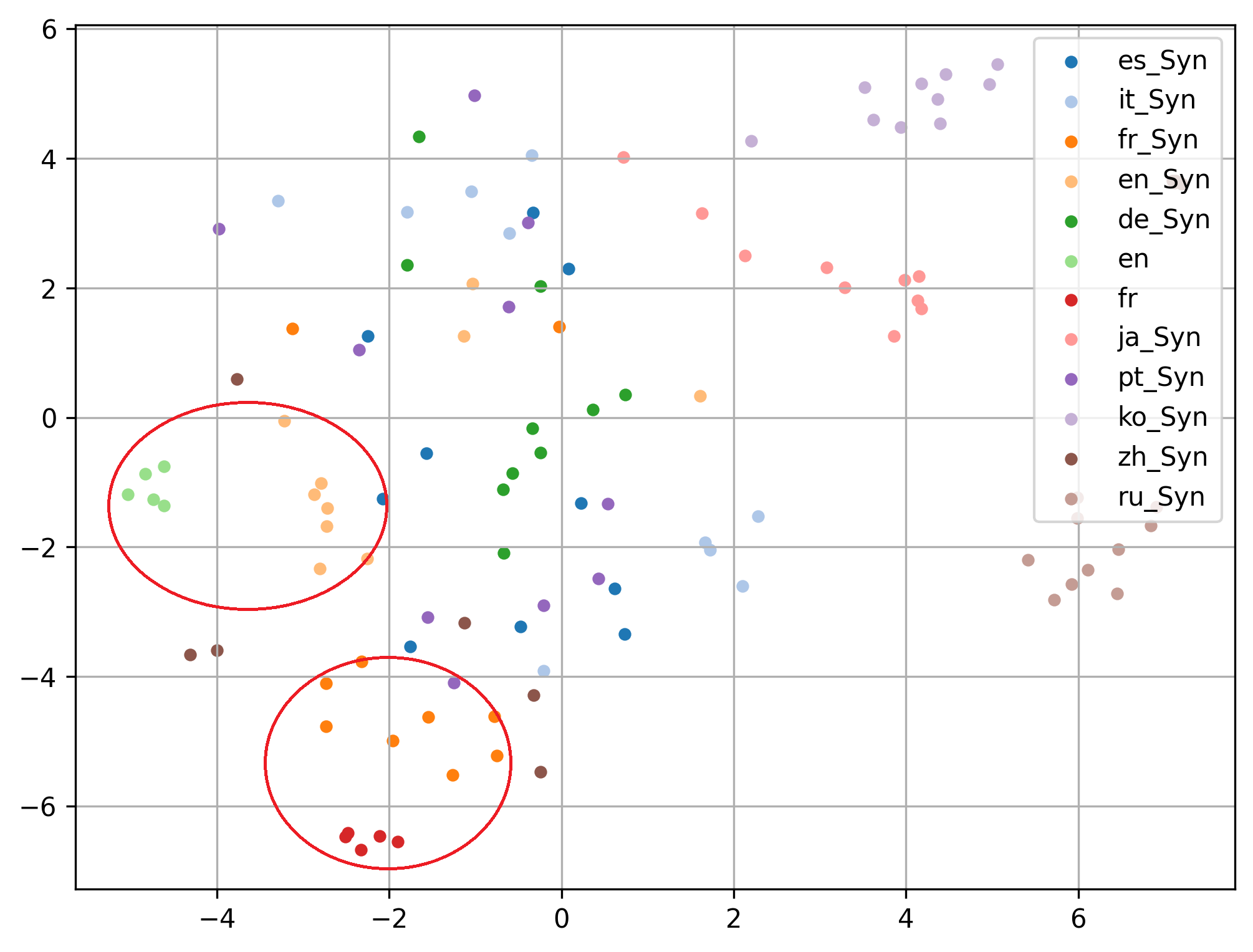}
  \caption{t-SNE Visualization of Real and Synthetic Speech Embeddings for Speaker $id011337$.}
  \label{fig:tsne}
\end{figure}

Fig.~\ref{fig:tsne} shows the t-SNE visualization of real and synthetic speech embeddings for speaker $id011337$. The synthetic embeddings are highly consistent with the real ones, indicating that Qwen3-TTS effectively preserves speaker identity. In particular, synthetic embeddings with the same language as the real speech (highlighted in red) are located very close to the corresponding real embeddings, while synthetic speech from other languages also forms well-clustered embeddings with clear separation. However, as shown in Table 1, augmenting the training data with synthetic speech does not lead to further performance improvements. Interestingly, training only on synthetic data achieves 1.022\% EER on tv26\_dev, which is close to the 0.95\% EER obtained with real data. It is noteworthy that these synthetic samples are generated using only about one-tenth of the real data as reference audio. This suggests that, under sufficient training data conditions, domain mismatch between synthetic and real data may degrade performance. In contrast, in low-resource scenarios, synthetic data augmentation can be a viable strategy to improve cross-lingual speaker verification.

\section{Conclusion}
This paper describes our SV systems for the TidyVoice2026 Challenge. In this work, we present a multilingual speaker verification system based on the large-scale self-supervised w2v-BERT 2.0 model, leveraging Layer Adapters and an MFA framework to extract robust speaker embeddings. To improve cross-lingual generalization, a language-adversarial training strategy using a GRL is introduced, encouraging the learning of language-invariant representations. Furthermore, we employ Qwen3-TTS, a multilingual ZS-TTS system, to synthesize additional speech for each speaker, enhancing language diversity in the training data. 
Experimental results on the TidyVoice 2026 development and evaluation sets demonstrate that fine-tuning on large-scale pre-trained models significantly improves speaker verification performance. Additionally, using SphereFace2 loss yields better results than ArcFace loss, while language-adversarial training provides modest improvements in suppressing language-specific variations, and synthetic data augmentation is effective under limited data conditions.

\section{Generative AI Use Disclosure}
Large Language Models (LLMs) were used exclusively for language editing, including rephrasing and grammatical refinement, to improve clarity and readability. The LLMs were not involved in the development of ideas, methodology design, experimental procedures, data analysis, or interpretation of results. All scientific content was developed and verified by the authors.
\bibliographystyle{IEEEtran}
\bibliography{mybib}

\begin{thebibliography}{10}
\providecommand{\url}[1]{#1}
\csname url@samestyle\endcsname
\providecommand{\newblock}{\relax}
\providecommand{\bibinfo}[2]{#2}
\providecommand{\BIBentrySTDinterwordspacing}{\spaceskip=0pt\relax}
\providecommand{\BIBentryALTinterwordstretchfactor}{4}
\providecommand{\BIBentryALTinterwordspacing}{\spaceskip=\fontdimen2\font plus
\BIBentryALTinterwordstretchfactor\fontdimen3\font minus \fontdimen4\font\relax}
\providecommand{\BIBforeignlanguage}[2]{{%
\expandafter\ifx\csname l@#1\endcsname\relax
\typeout{** WARNING: IEEEtran.bst: No hyphenation pattern has been}%
\typeout{** loaded for the language `#1'. Using the pattern for}%
\typeout{** the default language instead.}%
\else
\language=\csname l@#1\endcsname
\fi
#2}}
\providecommand{\BIBdecl}{\relax}
\BIBdecl

\bibitem{cai_resnet}
W.~Cai, J.~Chen, and M.~Li, ``Exploring the encoding layer and loss function in end-to-end speaker and language recognition system,'' in \emph{Proc. Odyssey}, 2018, pp. 74--81.

\bibitem{ecapa-tdnn}
B.~Desplanques, J.~Thienpondt, and K.~Demuynck, ``Ecapa-tdnn: Emphasized channel attention, propagation and aggregation in tdnn based speaker verification,'' in \emph{Proc. Interspeech}, 2020, pp. 3830--3834.

\bibitem{cam++}
H.~Wang, S.~Zheng, Y.~Chen, L.~Cheng, and Q.~Chen, ``Cam++: A fast and efficient network for speaker verification using context-aware masking,'' in \emph{Proc. Interspeech}, 2023, pp. 5301--5305.

\bibitem{redimnet}
I.~Yakovlev, R.~Makarov, A.~Balykin \emph{et~al.}, ``{Reshape Dimensions Network for Speaker Recognition},'' in \emph{{Proc. Interspeech}}, 2024, pp. 3235--3239.

\bibitem{wavlm}
S.~Chen, C.~Wang, Z.~Chen, Y.~Wu, S.~Liu, Z.~Chen, J.~Li, N.~Kanda, T.~Yoshioka, X.~Xiao \emph{et~al.}, ``Wavlm: Large-scale self-supervised pre-training for full stack speech processing,'' \emph{IEEE Journal of Selected Topics in Signal Processing}, vol.~16, no.~6, pp. 1505--1518, 2022.

\bibitem{wav2vec2.0}
A.~Baevski, Y.~Zhou, A.~Mohamed, and M.~Auli, ``wav2vec 2.0: A framework for self-supervised learning of speech representations,'' in \emph{Proc. NeurIPS}, 2020, pp. 12\,449--12\,460.

\bibitem{hubert}
W.-N. Hsu, B.~Bolte, Y.-H.~H. Tsai \emph{et~al.}, ``Hubert: Self-supervised speech representation learning by masked prediction of hidden units,'' \emph{IEEE/ACM transactions on audio, speech, and language processing}, vol.~29, pp. 3451--3460, 2021.

\bibitem{w2v-bert-2.0}
L.~Barrault, Y.-A. Chung, M.~C. Meglioli \emph{et~al.}, ``Seamless: Multilingual expressive and streaming speech translation,'' \emph{arXiv preprint arXiv:2312.05187}, 2023.

\bibitem{li2025enhancing}
Z.~Li, M.~Cheng, and M.~Li, ``Enhancing speaker verification with w2v-bert 2.0 and knowledge distillation guided structured pruning,'' in \emph{Proc. ICASSP}, 2026.

\bibitem{cai_asr_sv}
D.~Cai and M.~Li, ``Leveraging asr pretrained conformers for speaker verification through transfer learning and knowledge distillation,'' \emph{IEEE/ACM transactions on audio, speech, and language processing}, vol.~32, pp. 3532--3545, 2024.

\bibitem{mfa}
Y.~Zhang, Z.~Lv, H.~Wu, S.~Zhang, P.~Hu, Z.~Wu \emph{et~al.}, ``{MFA-Conformer: Multi-scale Feature Aggregation Conformer for Automatic Speaker Verification},'' in \emph{Proc. Interspeech}, 2022, pp. 306--310.

\bibitem{lora}
E.~J. Hu, Y.~Shen, P.~Wallis, Z.~Allen-Zhu, Y.~Li, S.~Wang \emph{et~al.}, ``Lora: Low-rank adaptation of large language models.'' in \emph{Proc. ICLR}, 2022.

\bibitem{cross-age}
X.~Qin, N.~Li, W.~Chao, D.~Su, and M.~Li, ``Cross-age speaker verification: Learning age-invariant speaker embeddings,'' in \emph{Proc. Interspeech}, 2022, pp. 1436--1440.

\bibitem{cosyvoice3}
Z.~Du, C.~Gao, Y.~Wang, F.~Yu, T.~Zhao, H.~Wang, X.~Lv, H.~Wang, C.~Ni, X.~Shi \emph{et~al.}, ``Cosyvoice 3: Towards in-the-wild speech generation via scaling-up and post-training,'' \emph{arXiv preprint arXiv:2505.17589}, 2025.

\bibitem{qwen3-tts}
H.~Hu, X.~Zhu, T.~He, D.~Guo, B.~Zhang, X.~Wang, Z.~Guo, Z.~Jiang, H.~Hao, Z.~Guo \emph{et~al.}, ``Qwen3-tts technical report,'' \emph{arXiv preprint arXiv:2601.15621}, 2026.

\bibitem{w2v-bert}
Y.-A. Chung, Y.~Zhang, W.~Han \emph{et~al.}, ``W2v-bert: Combining contrastive learning and masked language modeling for self-supervised speech pre-training,'' in \emph{Proc. ASRU}, 2021, pp. 244--250.

\bibitem{asp_pooling}
K.~Okabe, T.~Koshinaka, and K.~Shinoda, ``Attentive statistics pooling for deep speaker embedding,'' in \emph{Proc. Interspeech}, 2018, pp. 2252--2256.

\bibitem{GRL}
Y.~Ganin and V.~Lempitsky, ``Unsupervised domain adaptation by backpropagation,'' in \emph{Proc. ICML}, 2015, pp. 1180--1189.

\bibitem{libritts}
H.~Zen, V.~Dang, R.~Clark, Y.~Zhang, R.~J. Weiss, Y.~Jia, Z.~Chen, and Y.~Wu, ``{LibriTTS: A Corpus Derived from LibriSpeech for Text-to-Speech},'' in \emph{Proc. Interspeech}, 2019, pp. 1526--1530.

\bibitem{m2m100}
A.~Fan, S.~Bhosale, H.~Schwenk, Z.~Ma, A.~El-Kishky, S.~Goyal, M.~Baines, O.~Celebi, G.~Wenzek, V.~Chaudhary \emph{et~al.}, ``Beyond english-centric multilingual machine translation,'' \emph{Journal of Machine Learning Research}, vol.~22, no. 107, pp. 1--48, 2021.

\bibitem{whisper-large-v3}
A.~Radford, J.~W. Kim, T.~Xu, G.~Brockman, C.~McLeavey, and I.~Sutskever, ``Robust speech recognition via large-scale weak supervision,'' in \emph{Proc. ICML}, 2023, pp. 28\,492--28\,518.

\bibitem{tidyvoice2026}
A.~Farhadipour, J.~Marquenie, S.~Madikeri, T.~Vukovic, V.~Dellwo, K.~Reid, F.~M. Tyers, I.~Siegert, and E.~Chodroff, ``Tidyvoice 2026 challenge evaluation plan,'' \emph{arXiv preprint arXiv:2601.21960}, 2026.

\bibitem{vox2dev}
J.~Chung, A.~Nagrani, and A.~Zisserman, ``Voxceleb2: {Deep} {Speaker} {Recognition},'' in \emph{Proc. Interspeech}, 2018.

\bibitem{voxblink2}
Y.~Lin, M.~Cheng, F.~Zhang \emph{et~al.}, ``{VoxBlink2: A 100K+ Speaker Recognition Corpus and the Open-Set Speaker-Identification Benchmark},'' in \emph{{Proc. Interspeech}}, 2024, pp. 4263--4267.

\bibitem{3d-speaker}
S.~Zheng, L.~Cheng, Y.~Chen, H.~Wang, and Q.~Chen, ``3d-speaker: A large-scale multi-device, multi-distance, and multi-dialect corpus for speech representation disentanglement,'' \emph{arXiv preprint arXiv:2306.15354}, 2023.

\bibitem{kespeech}
Z.~Tang, D.~Wang, Y.~Xu, J.~Sun, X.~Lei, S.~Zhao, C.~Wen, X.~Tan, C.~Xie, S.~Zhou \emph{et~al.}, ``Kespeech: An open source speech dataset of mandarin and its eight subdialects,'' in \emph{Thirty-fifth conference on neural information processing systems datasets and benchmarks track (Round 2)}, 2021.

\bibitem{cnceleb1}
Y.~Fan, J.~Kang, L.~Li \emph{et~al.}, ``Cn-celeb: a challenging chinese speaker recognition dataset,'' in \emph{Proc. ICASSP}, 2020, pp. 7604--7608.

\bibitem{cnceleb2}
L.~Li, R.~Liu, J.~Kang \emph{et~al.}, ``Cn-celeb: multi-genre speaker recognition,'' \emph{Speech Communication}, vol. 137, pp. 77--91, 2022.

\bibitem{on-the-fly}
W.~{Cai}, J.~{Chen}, J.~{Zhang}, and M.~{Li}, ``{On-the-Fly Data Loader and Utterance-Level Aggregation for Speaker and Language Recognition},'' \emph{IEEE/ACM transactions on audio, speech, and language processing}, pp. 1038--1051, 2020.

\bibitem{musan}
D.~Snyder, G.~Chen, and D.~Povey, ``Musan: A music, speech, and noise corpus,'' \emph{arXiv preprint arXiv:1510.08484}, 2015.

\bibitem{RIR}
T.~Ko, V.~Peddinti, D.~Povey, M.~Seltzer, and S.~Khudanpur, ``A study on data augmentation of reverberant speech for robust speech recognition,'' in \emph{Proc. ICASSP}, 2017, pp. 5220--5224.

\bibitem{adamw}
I.~Loshchilov and F.~Hutter, ``Decoupled weight decay regularization,'' in \emph{Proc. ICLR}, 2019.

\bibitem{arcface}
J.~Deng, J.~Guo, N.~Xue, and S.~Zafeiriou, ``Arcface: Additive angular margin loss for deep face recognition,'' in \emph{{Proc. CVPR}}, 2019, pp. 4690--4699.

\bibitem{sf2}
B.~Han, Z.~Chen, and Y.~Qian, ``Exploring binary classification loss for speaker verification,'' in \emph{Proc. ICASSP}.\hskip 1em plus 0.5em minus 0.4em\relax IEEE, 2023, pp. 1--5.

\bibitem{qmf1}
J.~Thienpondt, B.~Desplanques, and K.~Demuynck, ``The idlab voxsrc-20 submission: Large margin fine-tuning and quality-aware score calibration in dnn based speaker verification,'' in \emph{Proc. ICASSP}.\hskip 1em plus 0.5em minus 0.4em\relax IEEE, 2021, pp. 5814--5818.

\bibitem{qmf}
Z.~Li, Y.~Lin, X.~Qin, N.~Jiang, G.~Zhao, and M.~Li, ``The dku-msxf speaker verification system for the voxceleb speaker recognition challenge 2023,'' \emph{arXiv preprint arXiv:2308.08766}, 2023.

\bibitem{simam}
X.~Qin, N.~Li, C.~Weng, D.~Su, and M.~Li, ``Simple attention module based speaker verification with iterative noisy label detection,'' in \emph{Proc. ICASSP}.\hskip 1em plus 0.5em minus 0.4em\relax IEEE, 2022, pp. 6722--6726.

\end{thebibliography}

\end{document}